# Long photoexcited carrier lifetime in a stable and earth-abundant zinc polyphosphide

Zhenkun Yuan,[1,2,3] Genevieve Amobi,[4] Shaham Quadir,[5] Smitakshi Goswami,[1,6] Guillermo L. Esparza,[7] Gideon Kassa,[1] Gayatri Viswanathan,[4,8] Joseph T. Race,[4,8] Muhammad R. Hasan,[4] Jack R. Palmer,[7] Sita Dugu,[5] Yagmur Coban,[1] Andriy Zakutayev,[5] Obadiah G. Reid,[5,9] David P. Fenning,[7] Kirill Kovnir,[4,8] Sage R. Bauers,[5] Jifeng Liu,[1] Geoffroy Hautier[1,2,3,*]

[1]Thayer School of Engineering, Dartmouth College, Hanover, NH, USA
[2]Department of Materials Science and NanoEngineering, Rice University, Houston, TX, USA
[3]Rice Advanced Materials Institute, Rice University, Houston, TX, USA
[4]Department of Chemistry, Iowa State University, Ames, IA, USA
[5]National Laboratory of the Rockies, U.S. Department of Energy, Golden, CO, USA
[6]Department of Physics and Astronomy, Dartmouth College, Hanover, NH, USA
[7]Aiiso Yufeng Li Family Department of Chemical and Nano Engineering, University of California San Diego, La Jolla, CA, USA
[8]Ames National Laboratory, U.S. Department of Energy, Ames, IA, USA
[9]Renewable and Sustainable Energy Institute, University of Colorado Boulder, Boulder, CO, USA
*Corresponding author. Email: geoffroy.hautier@rice.edu

## ABSTRACT

Halide perovskites have revolutionized optoelectronics by demonstrating that long carrier lifetime can be achieved in materials processed in relatively uncontrolled environments, whereas conventional inorganic semiconductors typically suffer from short carrier lifetime unless very carefully prepared and postprocessed. Here, we report the discovery of exceptionally long photoexcited carrier lifetime in monoclinic $ZnP_2$, effectively bridging the carrier lifetime gap between direct-gap inorganic semiconductors and halide perovskites. Through computational screening, $ZnP_2$ is identified as a long carrier lifetime semiconductor characterized by an unconventional polyphosphide bonding, combining covalently bonded phosphorus chains and polar-covalent Zn-P tetrahedra. Experimentally, $ZnP_2$ crystals synthesized from low-purity precursors exhibit bright band-to-band photoluminescence at 1.49 eV and carrier lifetimes of up to 1 μs. Further analysis reveals that the polyphosphide bonding of $ZnP_2$ suppresses the formation of deep intrinsic defects, making it defect resistant. Combined with its remarkable environmental stability, $ZnP_2$ presents a highly promising material for solar absorbers and light emitters. Our work illustrates that underexplored inorganic materials spaces with unusual chemical bonding hold great promise for discovering novel optoelectronic materials.

## INTRODUCTION

When electrons and holes are photoexcited or injected in a semiconductor, these excess carriers have a finite lifetime before recombination. The carrier lifetime is critically important for many optoelectronic devices: it dictates the efficiency of solar cells and light-emitting diodes (LEDs) as well as the sensitivity of photodetectors.[1-4] Achieving a long carrier lifetime is essential for these devices to deliver exceptional performance. The carrier lifetime is often limited by nonradiative recombination, which can cause significant carrier loss by dissipating photoexcited or injected carriers' energy as heat. Understanding and controlling nonradiative carrier lifetime in materials is central to the design and development of optoelectronic materials.[5-12]

The importance of long carrier lifetime has been highlighted in the last decade by the development of halide perovskites such as $CH_3NH_3PbI_3$ (MAPI).[13] Among direct-gap semiconductors, halide perovskites exhibit the longest carrier lifetime from initial reports of around 10 ns[14,15] to the more recent exceptional values exceeding 10 μs[16-18] (Figure 1). Combined with adequate band gap for solar absorption and sufficient carrier mobility,[19] the long carrier lifetime of halide perovskites is the foundation of their high performance as absorber layers in solar cells (reaching >27% efficiency in single-junction cells),[20] as well as the reason for their consideration in a widening range of optoelectronic applications, such as LEDs, lasers, and photodetectors.[13]

For visible light absorption and at moderate carrier densities, the carrier lifetime in the bulk of a semiconductor material is overwhelmingly controlled by the defect-assisted nonradiative recombination: deep defects capture electrons and holes through multiphonon emission, i.e., the Shockley-Read-Hall (SRH) process.[21,22] In this context, the long carrier lifetime and high performance of halide perovskites, even when grown in polycrystalline form in relatively uncontrolled environments (e.g., solution processing), have been attributed to their intrinsic high defect-tolerance.[7,23-27] However, despite their excellent optoelectronic properties and earth-abundance, halide perovskites face important challenges hindering their industrial deployment, among them their intrinsic instability (exhibiting e.g. air and moisture sensitivity).[28]

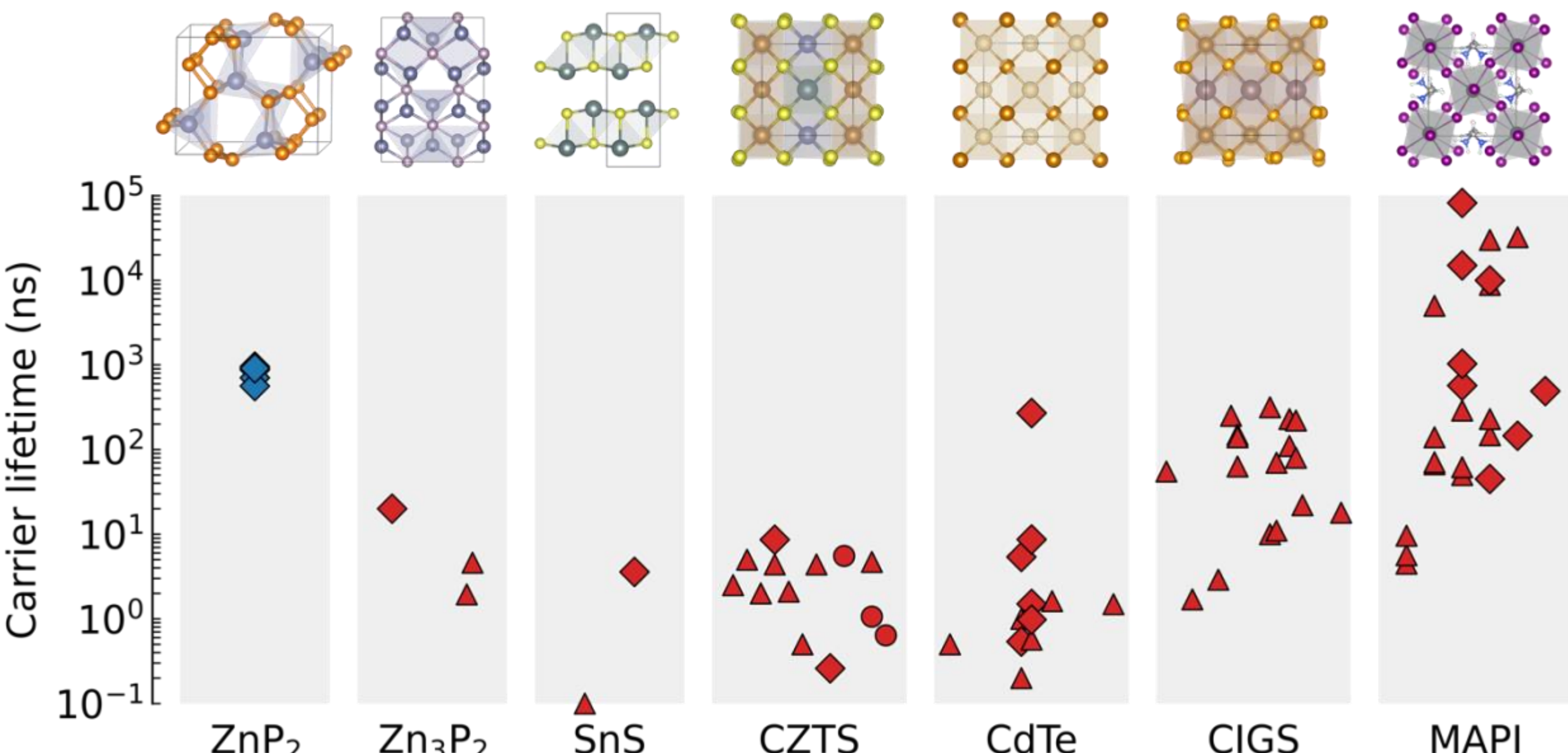

**Figure 1. Comparison of measured carrier lifetimes in $ZnP_2$ and well-studied photovoltaic absorber materials.** All the materials are direct-gap semiconductors. Data for $ZnP_2$ are first reported in this work, while those for all other materials were collected via a literature survey (see Tables S8–S13 for details). The diamond, triangle, and circle represent bulk crystal, thin film, and powder samples, respectively. All the data are for bare absorbers or equivalent, and were obtained by a range of characterization techniques.

In contrast to halide perovskites, conventional direct-gap inorganic semiconductors lag significantly behind in carrier lifetime, especially when used in polycrystalline form (Figure 1), but have a strong track record in long-term stability and deployment in commercial optoelectronic devices. Prominent examples include CdTe[29] and Cu(In,Ga)$Se_2$ (CIGS),[30] which are well-established thin-film solar absorbers, yet the carrier lifetime in these polycrystalline thin films is typically limited to ~1–200 ns.[5] The challenge in achieving long carrier lifetime in inorganic semiconductors is especially acute when earth-abundance is added as a constraint. $Zn_3P_2$,[31] SnS,[32] and $Cu_2ZnSn(S,Se)_4$ (CZTS),[33] which are widely studied earth-abundant absorbers for thin-film solar cells, have hardly reached carrier lifetimes higher than 10 ns (Figure 1).

It remains an open question whether inorganic semiconductors can achieve the favorable defect chemistry (e.g., high defect-tolerance) and long carrier lifetime of halide perovskites. Prime inorganic optoelectronic materials typically exhibit polar-covalent bonding often in zinc blende-related structures. Halide perovskites are significantly more ionic, and this has been used to rationalize their defect-tolerance.[27,34,35] While the chemical origin of the high defect-tolerance of halide perovskites is still under debate,[24] their success suggests that looking beyond the chemistries and bonding of traditional polar-covalent compounds could lead to the identification of long carrier lifetime inorganic optoelectronic materials.[36-39] To this end, high-throughput computational screening with virtually no constraints on chemical composition and structure provides a powerful tool to search outside traditional materials design spaces, especially when combined with follow-up experiments.[36,40,41]

Here, we bridge the carrier lifetime gap between direct-gap inorganic semiconductors and halide perovskites by reporting exceptionally long carrier lifetime in a zinc polyphosphide. We identify from computational screening the monoclinic phase of $ZnP_2$ as a promising long carrier lifetime direct-gap semiconductor. Unlike conventional inorganic semiconductors, $ZnP_2$ is characterized by covalently bonded phosphorus chains coexisting with polar-covalent Zn-P tetrahedra. $ZnP_2$ crystals exhibit bright band-to-band photoluminescence (PL) at 1.49 eV, and show long carrier lifetimes ranging from 500 ns up to 1 μs (Figure 1), measured by a variety of techniques including time-resolved photoluminescence (TRPL), time-resolved microwave conductivity (TRMC), phase fluorometry, and DC photoconductive current decay. The measured carrier lifetimes are remarkable for an unoptimized new inorganic material prepared from low-purity precursors (98.9–99.9% purity). We find that the long carrier lifetime of $ZnP_2$ is driven by its polyphosphide bonding, which suppresses the formation of deep intrinsic defects that are otherwise prevalent in $Zn_3P_2$. The combination of exceptional optoelectronic properties with environmental stability (in air, water, and acid) and earth-abundant constituents makes $ZnP_2$ of great interest for a variety of optoelectronic applications such as thin-film photovoltaics. More generally, our work illustrates the promise of discovering novel optoelectronic materials in underexplored inorganic materials spaces with unusual chemical bonding such as polypnictides.

## RESULTS

### Computational discovery of $ZnP_2$

Our discovery of the monoclinic $ZnP_2$ as a long carrier lifetime material starts from a computational screening that ranks candidate materials by bulk properties but also intrinsic point defects. The screening workflow is based on high-throughput Density Functional Theory (DFT) computations, and is similar to the one used in our previous work (see Methods).[36] In this work, we focus our screening on phosphides as they have shown attractive optoelectronic behavior.[36,42-49] We have screened approximately 1,400 known phosphides, focusing on those with suitable band gaps for optoelectronic applications in the visible and near-IR region (~0.9–2.5 eV) and with promising carrier lifetime. The carrier lifetime is obtained from a model including SRH recombination, using calculated formation energies and charge transition levels of intrinsic defects and assuming a constant yet reasonably large carrier capture coefficient ($10^{-6}$ $cm^3/s$).[50] In this way, candidate materials with a low concentration of deep intrinsic defects are predicted to have a long carrier lifetime. A series of phosphides stand out from the screening (Figure S1) with the monoclinic $ZnP_2$ being especially appealing as it combines favorable defect properties with earth-abundance and air stability (based on previous experimental synthesis reports).[51,52] Both Zn and P are non-critical elements, with an annual global production of 13.8 and 63.7 million tons, respectively.[53,54] We note that our screening indicates that $ZnP_2$ should significantly outperform $Zn_3P_2$ in terms of intrinsic defects and nonradiative carrier lifetime (Figures S1 and S2 and Note S1). $Zn_3P_2$ has been widely studied as an earth-abundant absorber for thin-film solar cells.[31]

$ZnP_2$ crystallizes in two polymorphs, tetragonal (space group $P4_32_12$) and monoclinic (space group $P2_1/c$).[51,52] Our interest is in the monoclinic phase (Figure 2A), which is also known as $\beta$-$ZnP_2$ or black $ZnP_2$.[51,52] Unless specifically mentioned, $ZnP_2$ in this work refers to the monoclinic phase. $ZnP_2$ is a polyphosphide in which both Zn and P are in a tetrahedral coordination environment and P atoms are arranged in semi-spiral chains parallel to the crystallographic *c*-axis.[51,52] Therefore, $ZnP_2$ is characterized by the coexistence of Zn-P tetrahedra with polar-covalent bonding (as in $Zn_3P_2$) and phosphorus chains with covalent bonding. The mixed nature of the bonding is illustrated by the electron localization function (ELF) analysis[55] in Figure 2A, which depicts the electron pair probability distribution for the Zn-P and P-P bonds. In $ZnP_2$, the formal oxidation state assignment is +2 for Zn (as in $Zn_3P_2$), while the presence of P-P bonds drives P to be formally $-1$, i.e., $Zn^{2+}(P^{1-})_2$. This assignment agrees with Bader charge analysis (Table S1). The electronic band structure of $ZnP_2$ exhibits a computed fundamental direct band gap of 1.46 eV and dispersive band edges indicative of relatively low effective mass for electrons (0.19–0.92 $m_0$) and holes (0.22–6.40 $m_0$) (Figure 2B and Table S2). The computed phonon-limited mobility ranges from 150 to 900 $cm^2/Vs$ for electrons and 20 to 750 $cm^2/Vs$ for holes (Table S2), and these values should be viewed as an upper bound as other scattering mechanisms (e.g., grain boundary and impurity scattering) could be present. Besides, the computed optical absorption coefficient is close to that of InP, a known phosphide semiconductor (Figure S3). The computed band gap, carrier mobility, and optical absorption of $ZnP_2$ are all adequate for a high-performance optoelectronic material, including use as absorber layers in thin-film solar cells.

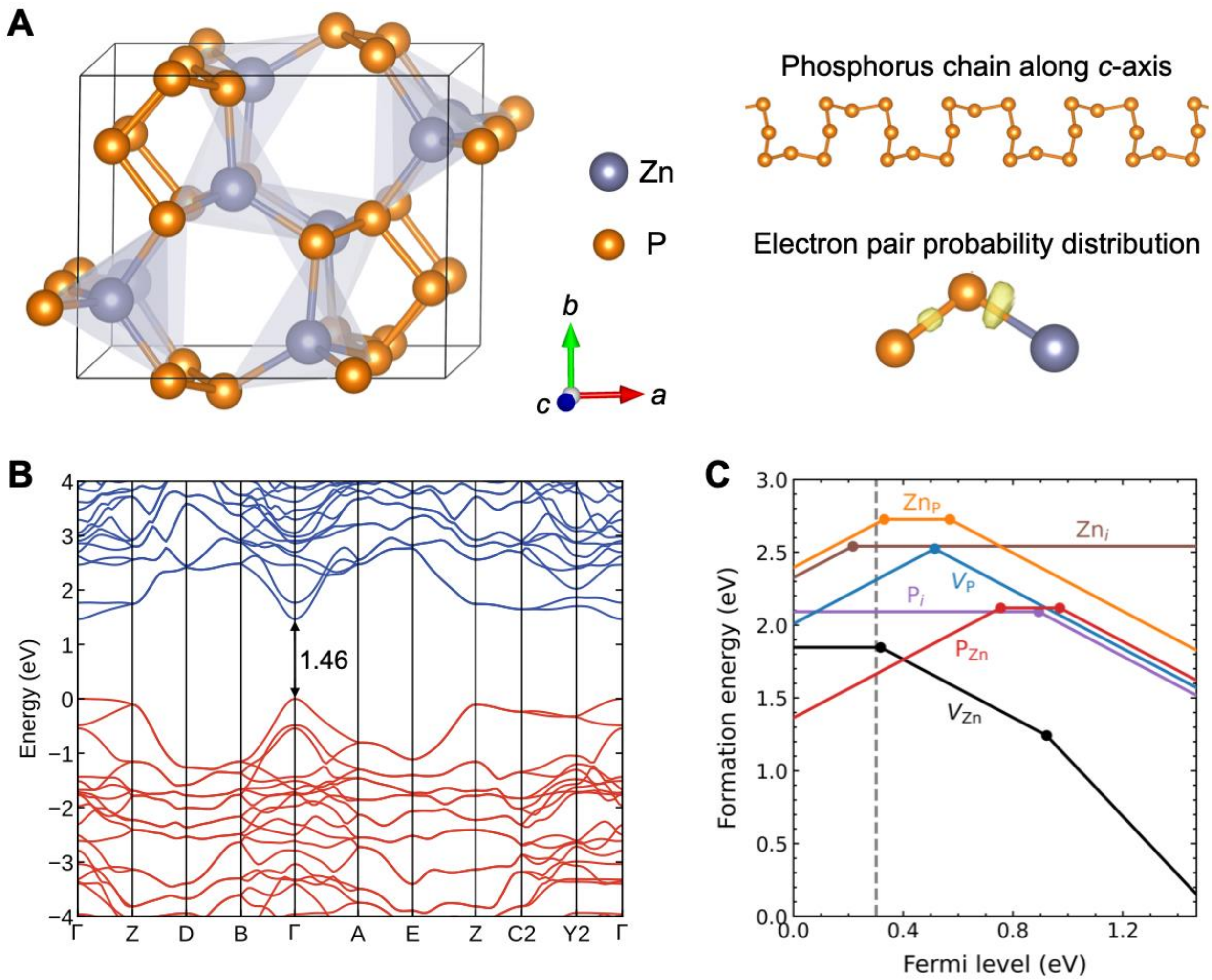

**Figure 2. Crystal structure and bonding, electronic band structure, and defect properties of $ZnP_2$.** (**A**) Crystal structure of monoclinic $ZnP_2$, with a highlight of the [$ZnP_4$] tetrahedra and the semi-spiral (infinite) phosphorus chains. Also displayed is the electron pair probability distribution for the Zn-P and P-P bonds as determined using the ELF (isosurface level 0.88). The ELF maps the electron pair probability distribution, allowing for a direct and intuitive analysis of chemical bonding.[55] (**B**) HSE06-calculated electronic band structure, with a 1.46 eV fundamental direct band gap. (**C**) HSE06-calculated formation energies as a function of Fermi level for the most important intrinsic point defects in $ZnP_2$. For each Fermi-level position, only the most stable charge state is plotted, with the dots (changes of slope) indicating changes in charge states (i.e., charge transition levels). The formation energies are for P-rich chemical-potential conditions (Figure S5). The gray dashed line indicates the equilibrium Fermi-level position, assuming synthesis of $ZnP_2$ at 1000 K followed by rapid quenching to 300 K and absence of any extrinsic dopants.

We have performed a systematic and more accurate calculation of intrinsic point defects in $ZnP_2$ using the hybrid functional (see Methods). Figure 2C shows formation energies for the most important intrinsic point defects in $ZnP_2$ (see Figure S4 for all defects). Here, we focus on the P-rich chemical potentials as our experimental syntheses reveal that these conditions are required to form the monoclinic phase (see discussion later); defect formation energies for other chemical-potential conditions are also provided in Figure S4. The formation energies determine how easily the defects will form during synthesis and the changes of slope in the formation-energy lines represent charge transition levels. Figure 2C indicates that the intrinsic defects will favor p-type doping, whereas for n-type doping, the zinc vacancies ($V_{\mathrm{Zn}}$), which are acceptors, will act as major electron killers. The gray dashed line in Figure 2C shows the Fermi level set by the intrinsic defects, assuming synthesis of $ZnP_2$ at 1000 K followed by rapid quenching to room temperature and absence of any extrinsic impurities. The calculated Fermi-level position indicates an intrinsic p-type material with a relatively low hole concentration ($\sim 10^{14}\ \mathrm{cm}^{-3}$; Table S3). Consistent with our defect calculations, all previous electrical measurements on as-grown $ZnP_2$ found p-type conductivity,[56-59] except for one study reporting n-type conductivity with gallium doping.[57]

The hybrid functional defect calculations confirm the potential to achieve long carrier lifetime in $ZnP_2$, corroborating the computational screening. As indicated in Figure 2C, while all intrinsic defects in $ZnP_2$ exhibit deep transition levels and therefore can potentially act as nonradiative recombination centers, they all have high formation energies (above $\sim$1.4 eV) for Fermi levels in intrinsic to p-type regions. The high formation energy will limit the concentration of the deep defects, reducing their contribution to nonradiative recombination. We have computed from first principles the nonradiative carrier capture coefficients for the two lowest-energy defects under P-rich conditions: $V_{\mathrm{Zn}}$ and P-on-Zn antisite ($\mathrm{P}_{\mathrm{Zn}}$). We find that the $V_{\mathrm{Zn}}$ will not be a strong nonradiative recombination center due to large hole and electron capture barriers (Figure S6). In contrast, the $\mathrm{P}_{\mathrm{Zn}}$ antisite is found to cause fast electron and hole capture with small energy barriers and large capture coefficients ($\sim 10^{-8}$–$10^{-6}$ $\mathrm{cm^3/s}$ at 300 K; Figures S7A–S7D). Assuming a high-temperature (1000 K) synthesis of $ZnP_2$, the $\mathrm{P}_{\mathrm{Zn}}$ concentration is estimated to be $10^{11}$–$10^{15}\ \mathrm{cm}^{-3}$ for Fermi level varying from the mid gap to 0.1 eV above the valence band (Figure S7E). As a result, the nonradiative carrier lifetime is predicted to range from 1 ns up to 100 μs (Figure S7E). These values are comparable to previous first-principles calculations with a similar level of theory on halide perovskites.[60] Our theoretical results suggest that the long carrier lifetime expected in $ZnP_2$ is a direct consequence of the high intrinsic defect formation energies, which arises from the polyphosphide bonding of this material (see Discussion section).

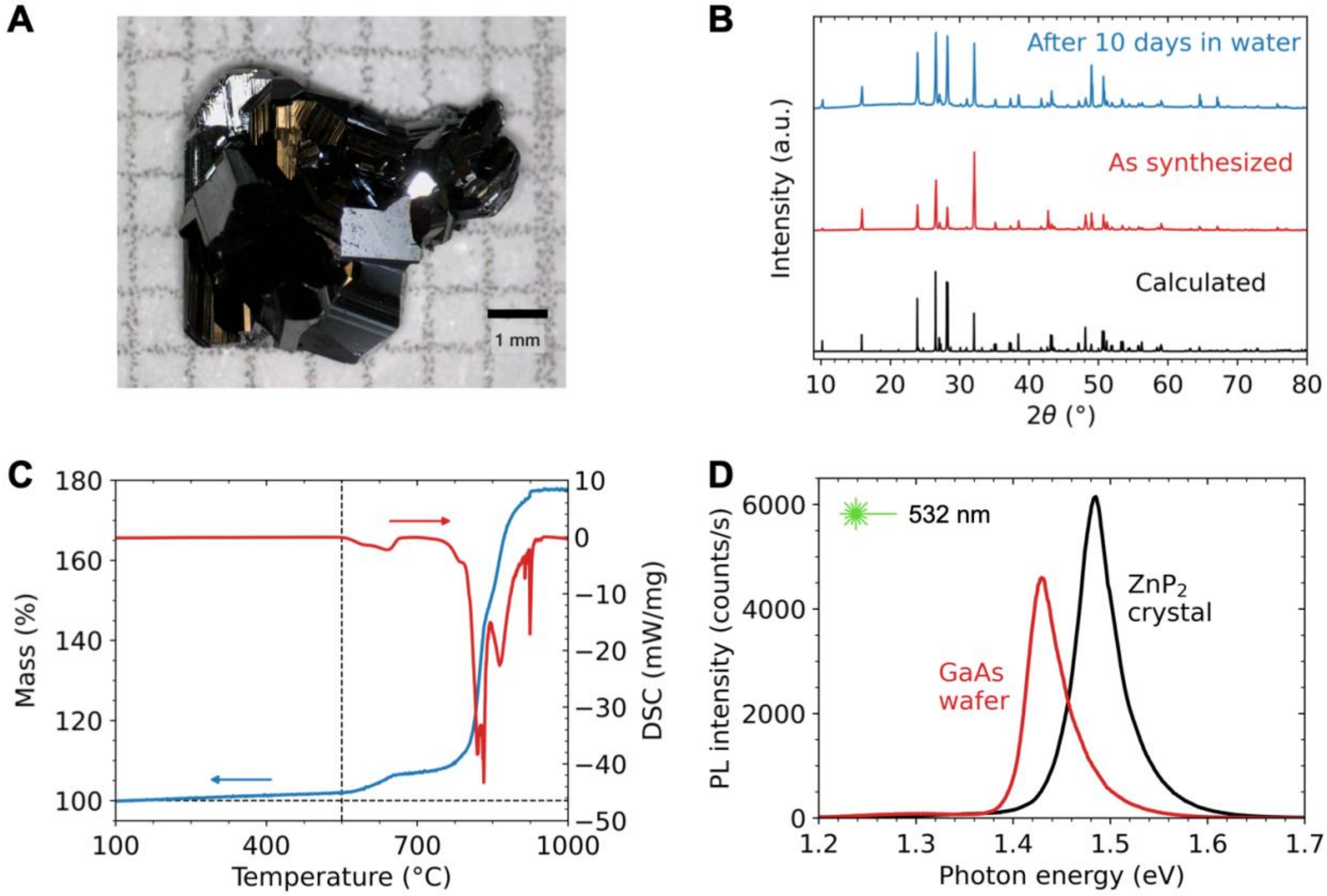

**Figure 3. Synthesized crystal, chemical stability, and PL.** (**A**) Image of an as-grown $ZnP_2$ crystal on millimeter-grid paper. (**B**) XRD pattern of the ground powder of an as-grown $ZnP_2$ crystal and of a (different) $ZnP_2$ crystal immersed in water for 10 days. The observed intensity changes are due to preferred orientation (see Figure S11 for Rietveld refinement). (**C**) TGA-DSC experiments in ambient air for the ground $ZnP_2$ crystal (the same sample labeled "As synthesized" in plot (B)). (**D**) Comparison of room-temperature PL spectra between an as-grown $ZnP_2$ crystal and a high-purity single-crystal GaAs wafer, both measured using 532 nm laser excitation and the same excitation power density.

### Synthesized crystals and stability of $ZnP_2$

We then move to experimental realization of $ZnP_2$ to verify our prediction of long carrier lifetime of this material. We have synthesized large millimeter-sized $ZnP_2$ crystals using a gas phase transport reaction at around 800 ℃ (see Methods). Figure 3A shows the image of an as-grown $ZnP_2$ crystal; images of other $ZnP_2$ crystals are found in Figure S8. The $ZnP_2$ samples appear black, indicating strong optical absorption. They are not single crystals, but rather each is an aggregate of small individual single crystals with different orientations. The key to obtaining this monoclinic phase and avoiding formation of the competing tetragonal phase (i.e., $\alpha$-$ZnP_2$) as well as $Zn_3P_2$ is using an excess of phosphorus during synthesis. The high phosphorus vapor pressure is critical for stabilizing the target phase, as found in previous work.[51,52] The absence of secondary phases is confirmed by powder X-ray diffraction (PXRD) performed on the ground $ZnP_2$ crystals (Figure 3B). Phase purity and composition of some of our samples have also been checked with Raman and EDS spectroscopy without the need to crush the crystals (Figures S9 and S10). We find that $ZnP_2$ is not sensitive to air. Our samples can be kept in ambient air for months without any visible degradation, in agreement with previous literature reports.[51,52] Even after immersion in water for 10 days, $ZnP_2$ shows no change in visual appearance, and the PXRD pattern retains all the peaks expected for the monoclinic phase, without any additional reflections corresponding to secondary phases (Figure 3B). Stability tests in concentrated acid (5N HCl) for up to 48 h also show no detectable degradation (Figure S12). Further, TGA-DSC experiments in ambient air find that $ZnP_2$ does not show appreciable oxidation until ~500 ℃ (Figure 3C). Together, these stability tests indicate that $ZnP_2$ is an environmentally stable material.

### Measured optoelectronic properties of $ZnP_2$

Moving to optoelectronic properties, Figure 3D displays the room-temperature photoluminescence (PL) spectra of an as-grown $ZnP_2$ crystal, showing an intense and narrow PL peak at 1.49 eV. The PL peak position agrees well with the calculated band gap (1.46 eV) as well as the step in the optical transmittance spectra (Figure S13), and can be attributed to band-to-band radiative recombination. In addition, the PL peak has a full width at half maximum (FWHM) of ~0.76×1.8 $k_{\mathrm{B}}T$ at 300 K, further suggesting that it is a single band-to-band transition.[61] The strong band-to-band PL emission indicates a highly luminescent material and implies weak nonradiative recombination. The bright room-temperature PL is atypical for an inorganic semiconductor grown without efforts at optimization.[49,62-65] The band-to-band PL intensity is comparable to that of a high-purity single-crystal GaAs wafer (Figure 3D). Additionally, we find that the same sample shows only a mild decrease in PL intensity after it has been in ambient air for about 5 months (Figure S14), in line with the chemical stability.

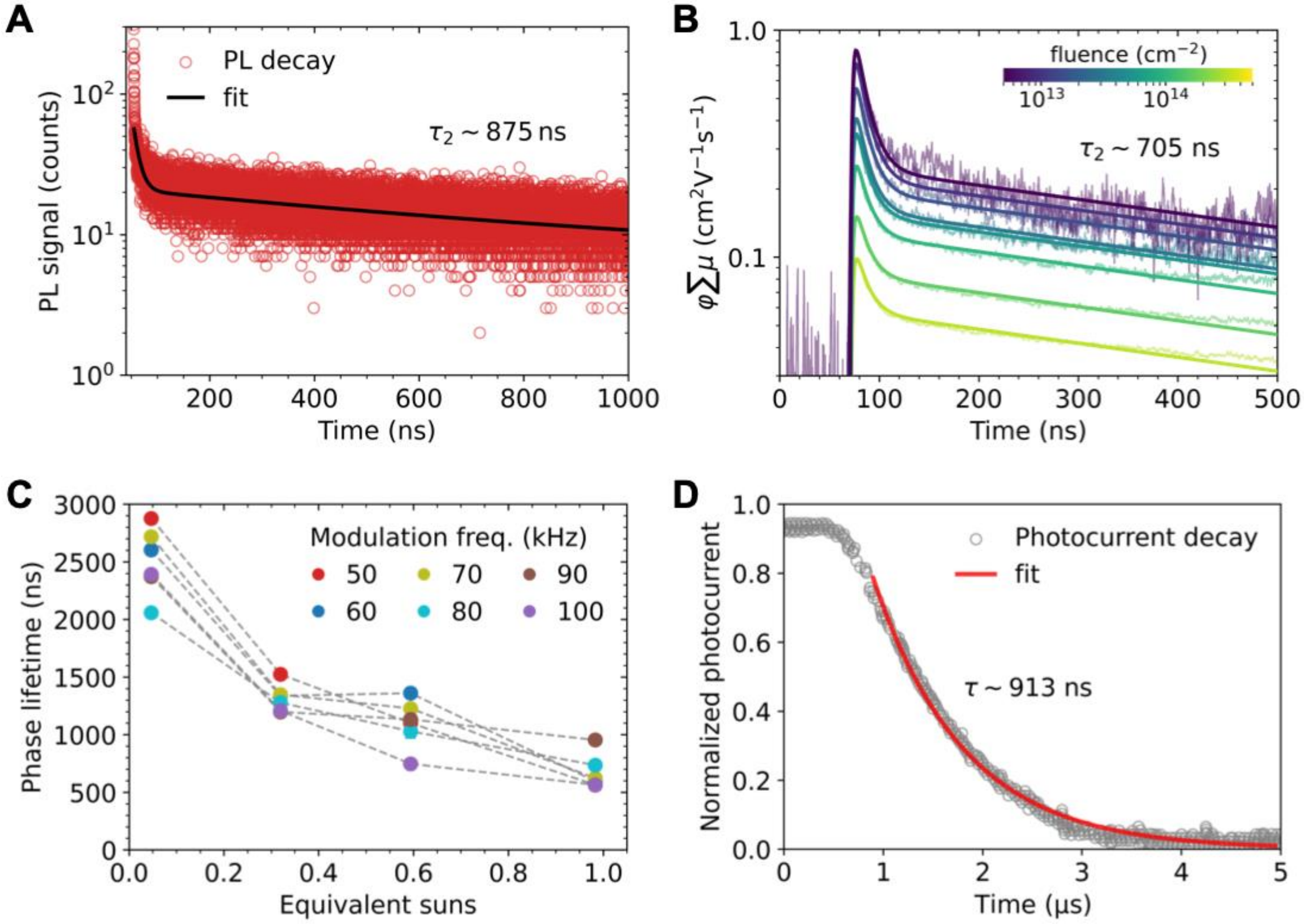

**Figure 4. Carrier lifetime measurements.** (**A**) TRPL decays of an as-grown $ZnP_2$ crystal, measured using 670 nm laser excitation, fluence $3.16 \times 10^{11}$ photons/cm$^2$/pulse, and repetition rate 125 kHz. (**B**) Fluence-dependent TRMC transients, measured using 532 nm laser excitation. (**C**) Phase fluorometry lifetime measured using 638 nm laser excitation, at different modulation frequencies. The emission was collected through a set of filters with an 800–875 nm bandpass, targeting the band-to-band transition. (**D**) DC photoconductive current decay, measured using 780 nm laser excitation and repetition rate 111 kHz.

We have experimentally evaluated the carrier lifetime of $ZnP_2$ using a variety of techniques (see Methods). Figure 4A shows the time-resolved PL (TRPL) decays of an as-grown $ZnP_2$ crystal measured using 670 nm laser excitation with a low fluence ($3.16 \times 10^{11}$ photons/cm$^2$/pulse) and repetition rate (125 kHz). The TRPL signals exhibit an initial fast drop followed by a much slower process and can be described by a biexponential decay. Similar TRPL decay behavior has been observed in perovskite crystals, where the fast and slow processes are often attributed to surface and bulk recombination, respectively.[66,67] Lifetime for the slow process is extracted to be 875 ns. This is a remarkably high value for an inorganic material, especially considering that our samples are not optimized and produced from conventional solid-state synthesis using low-purity precursors. A slightly higher fluence produces quite similar TRPL decays and lifetimes (Figure S15 and Table S5), suggesting that we have not saturated any traps that may exist.[68] We have also performed time-resolved microwave conductivity (TRMC) measurements that probe the change in electrical conductivity between dark and after light illumination and thus the mobility and lifetime of photoexcited carriers.[69,70] Figure 4B shows the TRMC transients, acquired with 532 nm laser excitation and much higher fluences (~2–3 orders of magnitude higher than the TRPL fluence). A global biexponential fit to the fluence dependent TRMC data reveals two decay lifetimes: $\tau_1 = 12$ ns and $\tau_2 = 705$ ns. The longer lifetime is consistent with the slow process of the TRPL decays. A final optical method we used is phase fluorometry, which is based on the measurement of phase shift and demodulation of the fluorescence emission of the sample under constant illumination to set the baseline injection condition topped with a sinusoidally modulated excitation light.[71] At 638 nm laser excitation and ~1-sun equivalent illumination, carrier lifetimes of ~550–950 ns depending on modulation frequency are obtained (Figure 4C), again indicative of long carrier lifetime. Therefore, all three optical techniques (TRPL, TRMC, and phase fluorometry) consistently reveal long carrier lifetime in our $ZnP_2$ samples, ranging from 500 ns to nearly 1 μs.

We have also measured the carrier lifetime based on DC photoconductive current decay, by making Ohmic contact to a $ZnP_2$ crystal using indium. We find that $ZnP_2$ is photoconductive; under illumination using a 780 nm laser diode at a power density of 0.164 W/cm$^2$, the sample exhibits a photoconductive current ~1000 times greater than the dark current (Figure S16). By following the photoconductive current decay with a monoexponential fit (Figure 4D), we obtain a 913 ns carrier lifetime, further supporting the long carrier lifetime found by the contactless optical methods.

Our results demonstrate exceptionally long carrier lifetime of $ZnP_2$, significantly surpassing those of well-established direct-gap inorganic semiconductors (e.g., CdTe and CIGS) and emerging ones (e.g., CZTS) (Figure 1). Notably, our characterization across TRPL, TRMC, phase fluorometry, and DC photoconductive current decay consistently yields long carrier lifetimes. This is remarkable because these techniques can sometimes produce different carrier lifetime results, as they can be dominated by different recombination mechanisms.[72-75] We note that our carrier lifetime characterization involves different $ZnP_2$ crystals prepared from different synthesis routes (Table S4). Additionally, for both TRMC and phase fluorometry measurements, we find that at-gap excitation wavelengths (~800 nm) lead to significantly reduced carrier lifetimes (though still >200 ns; Figures S17 and S18B). This contradicts the expectation that the measured carrier lifetime should be longer for longer excitation wavelengths as they penetrate deeper into the sample. However, we note that the $ZnP_2$ samples are not single crystals. Another explanation is that perhaps there exist near-surface traps capturing one type of the photoexcited carriers, leaving the other mobile with prolonged lifetime. We will address the surface chemistry of $ZnP_2$ in future studies.

In addition, we have examined the PL quantum yield (PLQY) for a $ZnP_2$ crystal. The PLQY gives a direct measure of the relation between nonradiative and radiative recombination.[5] Using ~1-sun equivalent illumination, we obtain a PLQY of $0.018\pm0.008\%$, based on the de Mello integrating sphere method.[76] This result surpasses the 0.01% threshold considered highly promising for further development of new optoelectronic material,[5] and exceeds the values obtained for CdTe thin films ($10^{-5}-10^{-4}$%) and CZTS crystals (0.001%).[77,78]

Finally, we find that $ZnP_2$ shows also promising carrier transport. From the 532 nm TRMC data, the sum of the photogenerated electron and hole mobilities is extracted to be ~1 $cm^2/Vs$ (Figure 4B), which is comparable to the TRMC sum mobility in some perovskite polycrystalline films.[72,79] The 800 nm TRMC yields a similar sum mobility of ~1–3 $cm^2/Vs$. The TRMC mobility is adequate for solar cell applications but much lower than the theoretically predicted averaged phonon-limited electron and hole mobility (~300 $cm^2/Vs$). Since our $ZnP_2$ samples are not single crystals nor optimized, additional scattering mechanisms (e.g., grain boundary and impurity scattering) could limit the carrier mobility. We have also performed DC Hall effect measurements on a $ZnP_2$ crystal. The sample shows p-type conductivity, with a moderate hole concentration of $10^{15}$ $\mathrm{cm}^{-3}$ and hole mobility of ~5 $cm^2/Vs$ at room temperature (Figure S19). The observed p-type character agrees with previous experiments[56-59] and defect calculations. We note that a previous photo-Hall effect study reported hole and electron mobility to be 35 and 81 $cm^2/Vs$ in as-grown p-type and Ga-doped (n-type) $ZnP_2$ crystals, respectively.[57]

## DISCUSSION

The exceptional optoelectronic behavior of the $ZnP_2$ is especially intriguing as $Zn_3P_2$ has long been considered a promising contender for earth-abundant solar absorber. $Zn_3P_2$ exhibits a fundamental direct band gap of 1.5 eV and adequate carrier mobility, which are quite similar to those of $ZnP_2$ (Table S2). However, despite decades of materials optimization and growth of high-quality thin films, $Zn_3P_2$ has only achieved carrier lifetime of at most 20 ns in a polycrystalline wafer,[80] while the reported carrier lifetimes in epitaxial thin films remain below 5 ns.[81,82] These values are significantly lower than our $ZnP_2$ measurements.

Many factors can contribute to the distinct carrier lifetimes of $ZnP_2$ and $Zn_3P_2$, including intrinsic defects, unintentional impurities, and differences in surface recombination. A comparison of intrinsic defect formation energies calculated using the hybrid functional in these two compounds confirms that deep defects have overall higher formation energies in $ZnP_2$ than in $Zn_3P_2$ (Figures S20A and S20B). The higher deep-defect formation energies in $ZnP_2$ can be clearly seen in Figures 5A and 5B, which compares the formation energies of the major deep defects (i.e., $V_P$, $P_{Zn}$, and $Zn_P$) in $ZnP_2$ and $Zn_3P_2$ as a function of phosphorus chemical potential ($\mu_P$) for Fermi level at 0.3 eV above the materials' respective valence bands (representing moderate p-type doping). All of these defects exhibit large local atomic relaxations associated with changes in charge states, suggesting fast carrier capture (Table S6). We note that while $ZnP_2$ and $Zn_3P_2$ are stable in different regions of the Zn-P phase diagram, Figure 5A depicts the defect formation energies across the full chemical-potential range for a convenient comparison.

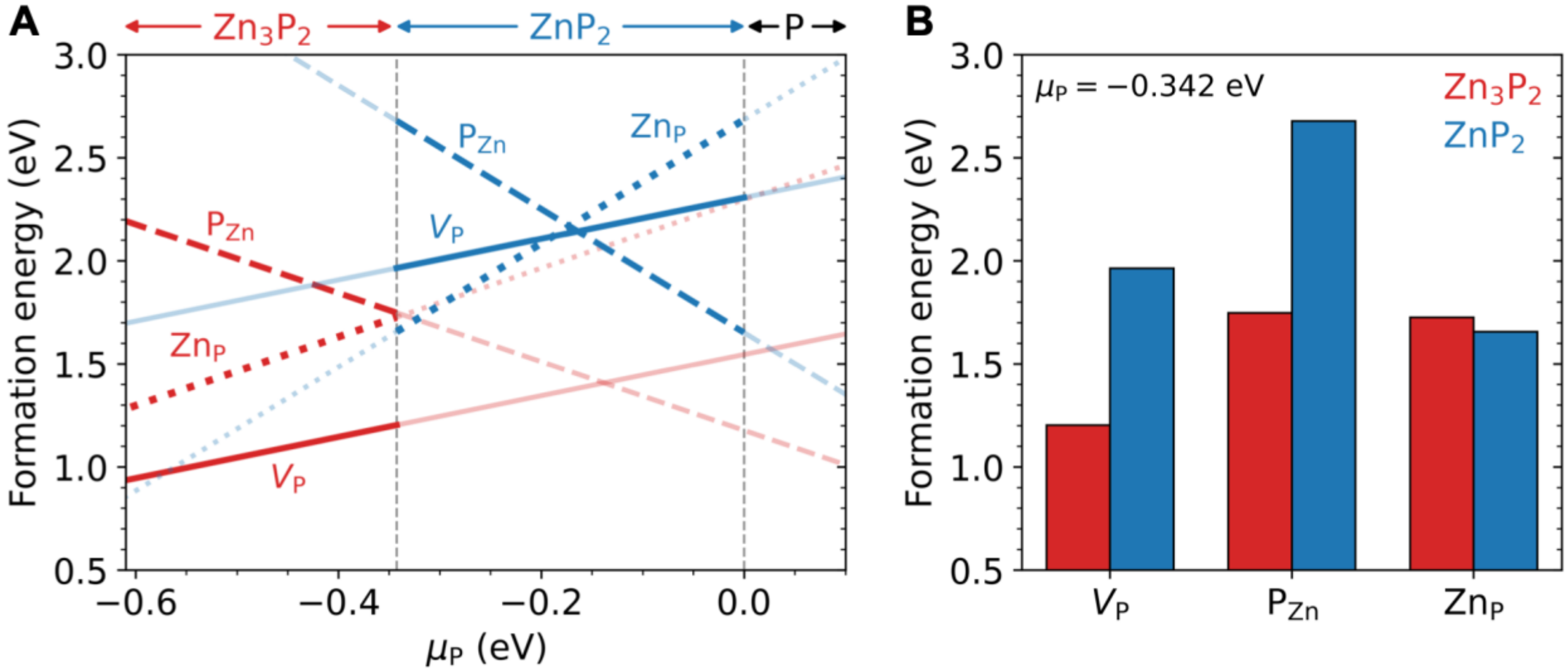

**Figure 5. Comparison of HSE06-calculated formation energies for the major deep intrinsic defects in $ZnP_2$ and $Zn_3P_2$.** (**A** and **B**) Formation energies (A) as a function of $\mu_P$ and (B) at a fixed $\mu_P$ value of $-0.342$ eV. The blue and red lines represent the formation energies of the defects in $ZnP_2$ and $Zn_3P_2$, respectively. These defect formation energies are obtained for Fermi level at 0.3 eV above the respective valence bands of $ZnP_2$ and $Zn_3P_2$. The vertical gray dashed lines indicate the thermodynamic stability boundaries: the first ($\mu_P = -0.342$ eV) is for $Zn_3P_2$ and $ZnP_2$, while the second ($\mu_P = 0$ eV) is for $ZnP_2$ and elemental phosphorus.

Figure 5A shows that the $V_P$ has a remarkably higher formation energy in $ZnP_2$ than in $Zn_3P_2$. This formation energy difference (>1 eV) is substantial, suggesting that significantly fewer deep phosphorus vacancies will be present in $ZnP_2$. The higher formation energy of $V_P$ in $ZnP_2$ is naturally first explained by its stability in a more P-rich chemical potential. However, bonding effects are also at play, as clearly indicated by the $V_P$ formation energies for the same $\mu_P$ value (Figure 5B). The formation of $V_P$ in $ZnP_2$ involves the breaking of strong covalent P-P bonds, which incurs a large energy cost. The polyphosphide bonding of $ZnP_2$ therefore directly drives its defect-resistance as it prevents the formation of deep phosphorus vacancies that are otherwise prevalent in the polar-covalent bonded $Zn_3P_2$. This is potentially a general tendency in polyphosphides. Indeed, the $V_{\mathrm{P}}$ in black phosphorus has a formation energy higher than 2.0 eV, and other transition metal polyphosphide in our dataset such as $CuP_2$ and $Re_2P_5$ also exhibit a high $V_{\mathrm{P}}$ formation energy (Figures S21 and S22).

Both the deep antisite defects $\mathrm{P_{Zn}}$ and $\mathrm{Zn_P}$ have also overall higher formation energies in $ZnP_2$ (Figure 5A). In particular, it is remarkable that $\mathrm{P_{Zn}}$, which should be energetically favored for high $\mu_{\mathrm{P}}$ necessary to stabilize $ZnP_2$, still exhibits a high formation energy. This can be understood from the fact that the electrostatics of replacing a $Zn^{2+}$ by a $P^{3+}$ or $P^{5+}$ (forming $\mathrm{P_{Zn}^{1+}}$ or $\mathrm{P_{Zn}^{3+}}$) is less favorable in $ZnP_2$ than in $Zn_3P_2$. This is because the antisite cation interacts with neighboring lattice $\mathrm{P^{1-}}$ in $ZnP_2$, whereas it interacts with lattice $\mathrm{P^{3-}}$ in $Zn_3P_2$. The preference of $\mathrm{P_{Zn}^{1+}}$ with $P^{3+}$ in $ZnP_2$, as opposed to $\mathrm{P_{Zn}^{3+}}$ with $P^{5+}$ in $Zn_3P_2$, is consistent with this rationale (Figures S20A and S20B). As it is the P-P bonds that drive the $-1$ oxidation state of P in $ZnP_2$, the high $\mathrm{P_{Zn}}$ formation energy is therefore related to the polyphosphide nature of this material.

## Conclusion

We have identified the zinc polyphosphide monoclinic $ZnP_2$ as a promising optoelectronic material that bridges the long-standing carrier lifetime gap between direct-gap inorganic semiconductors and halide perovskites. Through high-throughput computational screening and subsequent experimental verification, we have demonstrated that $ZnP_2$ can achieve exceptionally long carrier lifetime up to 1 μs in unoptimized crystals prepared from low-purity precursors. We have shown that the exceptional optoelectronic quality of $ZnP_2$ is intrinsically linked to its unconventional chemical bonding. The coexistence of covalently bonded phosphorus chains and polar-covalent Zn-P tetrahedra creates a defect resistant environment leading to high intrinsic defect formation energies. This effectively suppresses the formation of potential nonradiative recombination centers that have limited the carrier lifetime of related materials like $Zn_3P_2$. Furthermore, $ZnP_2$ addresses the critical stability concerns of halide perovskites. We have found remarkable environmental stability of $ZnP_2$ via months of air exposure, immersion in water for 10 days, and treatment with concentrated acid. Combined with its earth-abundant constituents and ~1.5 eV fundamental direct band gap, $ZnP_2$ presents a highly promising candidate for the next generation of stable, high-performance semiconductors, with potential use in solar photovoltaics, LEDs, and photoelectrochemical cells. Our work demonstrates that underexplored inorganic materials spaces with unusual chemical bonding, such as polypnictides, offer a fertile ground for discovering novel optoelectronic materials.

## METHODS

### High-throughput computational screening

The computational screening was based on high-throughput first-principles calculations using DFT, and the workflow is similar to the one designed in our previous work.[36] We started with a query of all the phosphide compounds in the Materials Project database (v2023.11.1).[83] We obtained ~1400 phosphide compounds by applying three filters to the query: (i) energy above hull within 0.1 eV, for a high likelihood of synthesizability; (ii) DFT band gap below 2.0 eV, targeting optoelectronic applications in the visible and near-IR region (and considering DFT band-gap underestimation); and (iii) excluding compounds containing H, C, N, or any elements of groups 16/17. Next, from electronic band structure calculations using the HSE06 hybrid functional,[84] we selected 92 phosphides predicted to have a fundamental direct band gap in the range ~ 0.9–2.5 eV. Besides the thermodynamic stability and band gap, we considered also intrinsic point defects. To this end, we performed first-principles defect calculations for each of the 92 phosphides. We ensured that the defect simulation cells (which are supercells) contain more than 200 atoms with lattice lengths >15 Å in all three directions, except for a few compounds with a large unit cell. A plane-wave energy cutoff of 420 eV and a $\Gamma$-only **k**-point grid were used in the defect calculations. We considered vacancy and antisite defects, and set the charge-state ranges for each defect based on the known oxidation states of the elements involved in creating the defect. The DFT band edges were corrected using the single-shot HSE06 method.[85] We successfully obtained defect results for 88 phosphides; the calculations for the other four phosphides (e.g., the f-electron compound $EuCd_2P_2$) have convergence issues. All the DFT calculations were performed using the PBE functional and PAW pseudopotentials as implemented in the VASP code.[86-89]

For each of the final 88 candidates, by obtaining the defect formation energies and charge transition levels, we identified the deep defects and computed key thermodynamic quantities including defect concentrations, Fermi-level position, and free carrier concentrations. Here, we assumed that defects with charge transition levels within the 25%–75% window of the band gap are deep. The Fermi-level position was obtained assuming a high-temperature (1000 K) synthesis followed by rapid quenching to room temperature and absence of any impurities. Using the calculated quantities, we computed the SRH electron-hole recombination rate ($R_{SRH}$) for each candidate material.[50] Two assumptions were made. First, we assumed a constant yet reasonably large value ($10^{-6}$ cm$^3$/s) for both the hole and electron capture coefficients for all deep defects. As such, the total $R_{SRH}$ rate of a material, which is the sum of the $R_{SRH}$ rates for all its deep defects, will be dominated by the deep defect having the lowest formation energy (i.e., highest concentration). Second, we assumed a photoexcited carrier density ($\Delta n$) of $10^{15}$ $\mathrm{cm}^{-3}$. Finally, the nonradiative carrier lifetime ($\tau_{\mathrm{SRH}}$) was determined by $\tau_{\mathrm{SRH}}=\Delta n/R_{SRH}$.[50] Since defect formation energies depend on elemental chemical potentials (representing material growth conditions), the carrier lifetime results for each candidate material fall in a certain range.

**First-principles study of $ZnP_2$**

First-principles calculations were performed for the electronic structure and intrinsic point defects in (monoclinic) $ZnP_2$, primarily based on the HSE06 hybrid functional and using tighter computational settings. All the calculations were performed using the VASP code with PAW pseudopotentials (PBE PAW datasets version 54). An energy cutoff of 400 eV was used for the plane-wave basis set.

HSE06 calculations were performed to determine the equilibrium crystal structure, electronic band structure, density of states (DOS), and optical absorption spectra of $ZnP_2$. Using a $4 \times 5 \times 5$ **k**-point grid for Brillouin-zone integration, the lattice constants of $ZnP_2$ were found to be: $a = 8.855$, $b = 7.295$, $c = 7.561$, and $\beta = 102.336°$, in good agreement with previous experiment.[52] For electronic band structure calculations, the high-symmetry paths were obtained following the crystallographic convention.[90] DOS was calculated using a $9 \times 10 \times 10$ **k**-point grid. For determination of the theoretical optical absorption spectra, the frequency-dependent dielectric function was calculated within the independent-particle approximation[91] and using a $7 \times 7 \times 7$ **k**-point grid; the real part of the dielectric function is derived from the imaginary part via the Kramers-Kronig transformation, using a small complex shift parameter of $10^{-5}$. Bader charge analysis,[92] electron localization function analysis,[55] and crystal orbital Hamilton populations (COHP) analysis[93,94] were also performed to understand the chemical bonding in $ZnP_2$. The DOS and COHP plots are found in Figure S23.

The electron and hole conductivity effective masses[95] were computed based on the HSE06-calculated electronic band structure on a uniform **k**-point grid (as for the DOS), assuming a doping level of $10^{16}$ $\mathrm{cm}^{-3}$. For the calculation of room-temperature phonon-limited carrier mobility, the AMSET approach[96] was used, assuming a doping level of $10^{16}$ $\mathrm{cm}^{-3}$ as well. The acoustic deformation potential scattering and polar optical phonon scattering were considered. All the material parameters for the AMSET inputs were calculated with the PBEsol functional[97] (Note S2).

Full HSE06 calculations were performed to study the intrinsic point defects in $ZnP_2$. A large supercell containing 432 atoms (when defect free) was used, derived from a $2 \times 3 \times 3$ repetition of the HSE06-relaxed unit cell. For the supercell containing a defect, a $\Gamma$-only **k**-point grid was used for Brillouin-zone integration, and all internal atomic positions were fully relaxed until the interatomic forces became smaller than 0.01 eV/Å. Spin polarization was included in all the defect calculations. Defect formation energies were computed using the standard formalism.[98] As an example, the formation energy of a phosphorus vacancy ($V_\mathrm{P}$) in the charge state $q$ (denoted as $V_\mathrm{P}^q$) is given by

$$E_f(V_P^q) = E_{tot}(V_P^q) - E_{tot}(\text{host}) + (\mu_P + E_P) + q(E_F + \text{VBM}), \quad (1)$$

where $E_{tot}(V_P^q)$ and $E_{tot}(\text{host})$ are the total energies of the $V_P^q$-containing and defect-free supercells, respectively. For charged defects, their supercell total energies were corrected,[99] using the calculated static dielectric constant. $E_F$ is the Fermi level referenced to the valence-band maximum (VBM). $\mu_P$ is the phosphorus chemical potential referenced to $E_P$ which is the energy of the P atom in elemental phosphorus solid. Higher $\mu_P$ values mean more P-rich conditions, thus lowering the $V_P$ formation energy, and vice versa. We chose $\mu_P = -0.342$ eV and 0 eV to represent P-poor and P-rich growth conditions, respectively (Figure S5 and Note S3). We investigated all possible intrinsic point defects in $ZnP_2$, including Zn vacancy ($V_{Zn}$), P vacancy ($V_P$), Zn-on-P antisite ($Zn_P$), P-on-Zn antisite ($P_{Zn}$), zinc interstitial ($Zn_i$), and phosphorus interstitial ($P_i$). There exist two and four inequivalent Zn and P lattice sites in the $ZnP_2$ unit cell, respectively, and this was considered in creating the vacancy and antisite defects. Care was exercised to correctly obtain the equilibrium atomic configurations of the defects.[100]

Nonradiative carrier capture at deep defects through multiphonon emission was studied based on a quantum mechanical one-dimensional (1D) model.[101] This started with constructing a 1D configuration coordinate diagram which depicts the potential energy surfaces of a defect in two adjacent charge states as a function of a generalized configuration coordinate ($Q$):

$$Q^2 = \sum_I M_I (\Delta \boldsymbol{R}_I)^2, \quad (2)$$

where $\Delta \boldsymbol{R}_I$ is the displacement of $I$-th atom with mass $M_I$. The atomic displacements are produced by a linear interpolation of the equilibrium configurations of the two charge states. For each charge state, this leads to a set of intermediate configurations, and the potential energy surface is obtained by calculating the total energies of these configurations and plotting the energies versus $Q$. The two potential energy curves are offset horizontally by $\Delta Q$, where $Q = \Delta Q$ and $Q = 0$ correspond to the equilibrium configurations of the initial and final charge states, and are offset vertically by $\Delta E$, where $\Delta E$ is the charge transition level referenced to the CBM (VBM) for electron (hole) capture.[101] The 1D configuration coordinate diagram forms the basis for analyzing carrier capture processes. For instance, one can analyze the electron and hole capture barriers, allowing for a qualitative estimate of how fast the capture will be.[101] To study nonradiative recombination of photoexcited electrons and holes via a defect level, a third potential energy surface is needed to represent an electron at the CBM and a hole at the VBM. This can be obtained by vertically shifting the potential energy curve of the final charge state by the band gap energy. In this work, using HSE06 hybrid functional, we computed the configuration coordinate diagram for $V_{Zn}$ and $P_{Zn}$.

The nonradiative carrier capture coefficients were calculated based on Fermi's golden rule and the static coupling approximation:[101,102]

$$C(T) = gV \frac{2\pi}{\hbar} \sum_m w_m \sum_n \left| \langle \chi_{im} | Q - Q_0 | \chi_{fn} \rangle \right|^2 |W|^2 \delta(E_{im} - E_{fn}), \quad (3)$$

where $g$ is the degeneracy factor of the final state, and $V$ the supercell volume. The evaluation of this equation involves two parts. The first is the vibronic overlap between the two potential energy surfaces, with $\chi_{im}$ and $\chi_{fn}$ being the vibrational wave functions for the $m$-th and $n$-th vibrational states of the initial $(i)$ and final $(f)$ charge states, respectively. $w_m$ is the thermal occupation of the vibrational state $m$, which sets temperature dependence. The second is the electron-phonon coupling. In the first-order perturbation theory, the electron-phonon coupling strength $|W|$ is given by $W = \langle \psi_b | \delta h / \delta Q | \psi_d \rangle$, where $\psi_b$ and $\psi_d$ are the bulk and defect states of the single-particle electronic Hamiltonian $h$ for the defect supercell, respectively. In this work, the $W$ elements were calculated using the HSE06 hybrid functional and the projector augmented wave formalism as implemented in the VASP.[102] For carrier capture by a charged defect, $|W|$ will be scaled by a Sommerfeld parameter, which captures the Coulomb interaction between the delocalized carrier and charged defect. The $\delta$-function ensures energy conservation, where $E_{im}$ and $E_{fn}$ are the total energies of the vibronic states; in the harmonic approximation, $E_{im} - E_{fn} = \Delta E + m\hbar\omega_i - n\hbar\omega_f$, where $\omega_i$ and $\omega_f$ are phonon frequencies of the initial and final potential energy surfaces, respectively.

Nonradiative carrier capture coefficients were obtained for the deep defect $\mathrm{P_{Zn}}$. In $ZnP_2$, the $\mathrm{P_{Zn}}$ is an amphoteric defect with two deep transition levels in the band gap: $(0/-)$ at 0.970 eV and $(0/+)$ at 0.755 eV (both referenced to the VBM). Correspondingly, this defect involves four capture processes: $C_n^0$ and $C_p^-$ for the $(0/-)$ transition, and $C_p^0$ and $C_n^+$ for the $(0/+)$ transition; the superscript and subscript denote the defect charge state and the type of carrier capture, respectively (Figure S7). For an amphoteric defect with $(0/-)$ and $(0/+)$ transition levels in the band gap, the SRH electron-hole recombination rate ($R_{SRH}$) is given by,[103]

$$R_{\mathrm{SRH}} = N_{\mathrm{T}}(np - n_0 p_0) \frac{C_n^+ C_p^0 P^- + C_n^0 C_p^- N^+}{N^+ P^- + P^0 P^- + N^+ N^0}, \quad (4)$$

where $N^+ = nC_n^+ + e_p^+$, $N^0 = nC_n^0 + e_p^0$, $P^0 = pC_p^0 + e_n^0$, and $P^- = pC_p^- + e_n^-$. $n = n_0 + \Delta n$ and $p = p_0 + \Delta n$, where $n_0$ and $p_0$ are the equilibrium electron and hole concentrations, respectively, while $\Delta n$ is the photoexcited carrier density. $e_p^+$, $e_p^0$, $e_n^0$, and $e_n^-$ are the emission coefficients; they are negligibly small for the $\mathrm{P_{Zn}}$ due to the deep defect levels. The $\mathrm{P_{Zn}}$ concentration was obtained by $N_{\mathrm{T}} = \sum_q N_{\mathrm{sites}} \exp(-E_{\mathrm{f}}(\mathrm{P}_{\mathrm{Zn}}^q)/k_B T)$, where the formation energy of $\mathrm{P}_{\mathrm{Zn}}^q$, $E_{\mathrm{f}}(\mathrm{P}_{\mathrm{Zn}}^q)$, depends on the Fermi level (as in Equation 1) and $N_{\mathrm{sites}}$ is the number of zinc sites (per unit volume) at which the antisite can form. To understand the doping effects, we considered different Fermi level positions which affect $R_{SRH}$ via $N_{\mathrm{T}}$ as well as $n_0$ and $p_0$. Finally, using the calculated $R_{SRH}$ rates of the $\mathrm{P_{Zn}}$ and assuming $\Delta n$ to be $10^{15}\ \mathrm{cm^{-3}}$, we determined the nonradiative carrier lifetime ($\tau_{\mathrm{SRH}}$) of $ZnP_2$ via $\tau_{\mathrm{SRH}} = \Delta n / R_{SRH}$.[50]

The open-source software, including PyCDT,[104] pymatgen-analysis-defects,[105] Pydefect,[106] py-sc-fermi,[107] Nonrad,[102] and CarrierCapture[108] were used in the defect calculations. The computer program VESTA[109] was used to visualize atomic structures, and the Python toolkit sumo[110] was used to plot the electronic band structure.

**Synthesis of $ZnP_2$**

The $ZnP_2$ crystals used in this work were obtained by four different synthetic routes involving either direct reactions between elemental zinc and phosphorus or reactions where potassium was present in the nominal mixture of the reactants. All reagents were used as received: Zn (powder, Alfa Aesar, purity level 99.9%) and red phosphorus (powder, Alfa Aesar, purity level 98.9%) as well as K (metal, Alfa Aesar, 99.95%) and Sn (shots, Alfa Aesar, 99.99+%). In all cases, formation of the monoclinic phase of $ZnP_2$ required high phosphorus vapor pressure. For potassium involving reactions, the presence of potassium increased the effective phosphorus vapor pressure, whereas in potassium-free reactions excess phosphorus was used to achieve the same effect.

Method 1 (K-Zn-P, two-step annealing): $ZnP_2$ crystals were obtained as byproducts of reactions targeting potassium zinc phosphide (KZnP) using elemental K, Zn, and P. The Zn:P ratio in the starting mixtures was approximately 1:1.4. Reactions were carried out in carbonized silica ampoules sealed under vacuum, with a two-step annealing. In the first annealing step, the reaction mixture was heated to 850 °C in 2 h and held at that temperature for 2 h, followed by natural cooling of the furnace. The resulting product was then ground, resealed in another silica ampoule, and annealed at 750°C for 72 h. After this second step, the targeted KZnP remained as a powder at the bottom of the ampoule, while well-formed $ZnP_2$ crystals were found grown on the walls or at the top of the ampoule. Because the ampoules were placed vertically in the furnace, crystallization of $ZnP_2$ at the top of the ampoule indicates formation via a vapor-phase transport mechanism. This might be induced by minute temperature gradient present between the furnace floor (where the elements are heated) and top of the ampoule (which is in the middle of the furnace chamber).

Method 2 (K-Zn-P, slow-cooling): A second profile using the same K, Zn, and P ratio as Method 1 also yielded $ZnP_2$ crystals. In this method, the reaction mixture was heated to 850 °C in 2 h and held for 2 h, then slowly cooled to 750 °C for over 48 h. The temperature was maintained at 750 °C for 24 h before allowing the furnace to cool naturally to room temperature. Similar to Method 1, $ZnP_2$ crystal grew on the walls at the top of the ampoule.

Method 3 (K-Zn-P, Sn-flux): $ZnP_2$ crystals were additionally obtained during attempts to grow single-crystal KZnP using Sn flux. In these reactions, the ratio of the combined reactants (elemental K, Zn, and P) to Sn flux was 1:80. The reaction mixture was heated to 850 °C for 10 h and held at that temperature for 24 h, followed by a slow cooling over 6 days to 350 °C, when the excess Sn was centrifuged off. Large $ZnP_2$ crystals were mechanically selected.

Method 4 (Zn-P with excess P): $ZnP_2$ crystals were also synthesized directly from elemental Zn and P in reactions containing excess phosphorus. Alumina crucibles were used to contain a mixture of Zn and P. The crucibles were placed in silica ampoules (14 mm ID,19 mm OD) which were evacuated. When stoichiometric Zn:P ratios were used, the tetragonal polymorph of $ZnP_2$ preferentially formed. To stabilize the monoclinic phase, excess phosphorus was added (Zn:P=1:2.41), and the temperature profile was similar to that of Method 2: heating to 850 °C for over 2 h with a 2 h dwell, followed by slow cooling to 750°C for over 48 h, and finally a 24 h dwell at 750 °C. When $ZnP_2$ crystals did not form, the obtained product was ground and annealed using the same heating profile. The $ZnP_2$ crystals formed on top of the alumina crucibles.

Our experimental efforts toward $ZnP_2$ crystals are summarized in Figure S8 and Table S4, where we show the sample images and detail their synthesis routes and characterization performed.

**Structural characterization**

Powder X-ray diffraction (PXRD) and thermogravimetric analysis/differential scanning calorimetry (TGA-DSC) measurements were performed on the ground powder of a $ZnP_2$ crystal. PXRD data were collected using a Rigaku MiniFlex 600 diffractometer equipped with Cu-$K_\alpha$ radiation and Ni-$K_\beta$ filter. The crystal was finely ground and dispersed onto a zero-background silicon sample holder. Diffraction patterns were recorded over a 2θ range from 3° to 80° with a scan speed of 10°/min and a step size of 0.2°.

TGA-DSC experiments for the ground $ZnP_2$ crystal were carried out in ambient air using a NETZSCH STA 449 F1 thermal analyzer. Measurements were performed from 40 to 1000 °C at a heating rate of 10 °C/min under an air flow of 20 mL/min using alumina crucibles.

Water stability tests were conducted on a $ZnP_2$ crystal after Hall measurements on this sample (see below). The crystal was immersed in water for 10 days, after which it was ground and analyzed by PXRD. In addition, the chemical stability of $ZnP_2$ was evaluated in concentrated HCl. An initial test was performed using a multi-phase (“dirty”) sample containing $KZn_4P_3$, $Zn_3P_2$, and monoclinic $ZnP_2$. PXRD analysis showed that all the secondary phases decomposed within 1 h, while $ZnP_2$ was left unchanged. In another test, a $ZnP_2$ crystal was kept in concentrated HCl for more than 48 h, after which the crystal was still intact (Figure S12).

Scanning Electron Microscopy and Energy Dispersive X-ray Spectroscopy (SEM-EDS) were used to study the surface morphology and elemental composition of the synthesized $ZnP_2$ crystals. The samples were examined using a TESCAN Vega 3 scanning electron microscope equipped with an EDS system. SEM imaging was conducted at an accelerating voltage of 30 kV under high-vacuum conditions. The crystal samples were mounted on aluminum stubs using carbon tape without any conductive coating, as the samples exhibited adequate surface conductivity to prevent charging during imaging. SEM was used to evaluate surface uniformity, grain morphology, and microstructural features in these samples. The SEM images were acquired at scale bar of 100 μm. EDS analysis was performed using an EDAX detector operated through APEX software. Elemental mapping and point analyses were performed on selected regions of the crystal surface to qualitatively and semi-quantitatively confirm the spatial distribution of constituent elements and rule out contamination or incorporation of K and Sn elements that were present in some synthetic methods.

**UV-Vis spectroscopy**

Diffuse reflectance spectroscopy measurements were conducted using a Perkin Elmer Lambda 1050+ UV/Vis/NIR spectrophotometer equipped with a 150 mm Spectralon-coated integrating sphere. Data were collected for the finely ground powder of a $ZnP_2$ crystal in the range 250–2500 nm. The sample was loaded in a spherical holder, pressed against a quartz lens, and then held in place with a spring. The sample holder was placed at the reflectance port while the specular port was open, and beam size was adjusted and focused on the sample only. A Spectralon reference standard was used as a blank. Diffuse reflectance data were converted to the corresponding absorption spectra using the Kubelka-Munk function.

**PL and Raman spectroscopy**

Photoluminescence (PL) and Raman spectroscopy of $ZnP_2$ crystals were measured using a Horiba LabRam micro-PL/Raman system under ambient conditions to investigate the optical and vibrational properties of the material. Instrument calibration was verified using a standard GaAs wafer as a reference. Both the PL and Raman spectra were collected across multiple spots of the samples, ensuring reproducibility and minimizing site-specific artifacts. For PL measurements, we used a 532 nm laser source with an output power of 100 mW attuned with a 0.01% Neutral Density (ND) filter. The optical efficiency of the equipment was determined to be around 28%, thereby yielding an effective laser power at the sample to be ~2.8 μW, which was sufficient to generate measurable PL. The laser spot size is 3 μm in diameter and thus a spot area of 7.07 $\mu m^2$. As a result, the effective excitation power density of the laser at a particular spot on the sample is 39.6 $W/cm^2$ corresponding to 396 suns of AM1.5G equivalent illumination. The PL spectra were collected in the range 540–1050 nm. Raman spectra were collected under the same optical configuration but attenuated with a 5% ND filter and the spectra were collected in the range 50–500 $\mathrm{cm}^{-1}$.

**TRPL measurements**

Time-resolved PL (TRPL) measurements were performed at room temperature using a PicoQuant 670 nm diode laser equipped with an 11 mm focal-length objective. The laser was operated at a repetition rate of 125 kHz using a driver with a variable repetition rate. The pulse width is 90 ps. Neutral density (ND) filters were placed in both the excitation and detection paths to adjust the photon fluence. The laser power density at the sample was measured to be 1 μW, with a beam spot size of ~50 μm in diameter. This corresponds to an excitation fluence of approximately $1.37 \times 10^{12}$ photons/cm$^2$/pulse. A 700 nm long-pass filter was placed in the optical path before the silicon avalanche photodiode to eliminate any laser light that had gotten through the dichroic mirror in the center cube. The TRPL decays were fitted (starting from 2 ns after the peak) using biexponential functions: $I(t) = A_1 \exp{(-t/\tau_1)} + A_2 \exp{(-t/\tau_2)} + C$, yielding two time constants $\tau_1$ and $\tau_2$ for the fast and slow processes, respectively.

**TRMC measurements**

Time-resolved microwave conductivity (TRMC) measurements were conducted as previously described.[70] Briefly, we mounted a $ZnP_2$ crystal on one of our standard 11×25×1 mm quartz substrates using paraffin wax as a microwave transparent adhesive. A small puddle (96 mg) of wax was created by heating the substrate with wax chips on it to ~50 ℃. The crystal was set in the puddle, which was then allowed to solidify. Optical excitation of the crystal was conducted from the front to avoid scattering losses from the wax. The crystal volume was calculated from its measured mass (157.6 mg) and theoretical density (3.54 g/cm$^3$). The crystal was approximated as a rectilinear volume with the same dimensions as the crystal major axes (7.3 × 4 × 2.7 mm) for the electromagnetic simulations needed to quantify the dark (i.e., equilibrium) conductivity and photoconductivity. The parameters that emerged from fitting the simulated cavity response to the experimental data were corrected by a factor of 1.77 to account for the difference between the actual crystal volume (44.5 mm$^3$) and the volume used in the simulation (78.8 mm$^3$). Simulations of the RF response of the microwave cavity as a function of the properties of the crystal were carried out as previously described,[70] using the RF module of COMSOL Multiphysics (version 6.3).

Both the equilibrium and time-resolved properties of the crystal were obtained from these measurements. The effective dark conductivity (0.019 S/cm) and relative permittivity (30) of the crystal were obtained by fitting the simulations to experimental resonance curves as well as by analyzing the difference between the slide with only wax and that with the crystal added. The sensitivity factor for calculating photoconductivity was obtained from the same fit, and was given by the partial derivative of the microwave power reflectance with respect to conductivity (−158.9).

**Phase fluorometry measurements**

Phase fluorometry measurements were performed using a home-built setup (see Figure S18A). The sample was illuminated with laser light consisting of a baseline offset (corresponding to the reported equivalent photon flux) with a 20% sinusoidal modulation. The laser diodes (638 nm and 785 nm 14-pin Butterfly Laser Diodes, Beijing RealLight Technologies Co., Ltd) are driven by a laser diode controller (LDC502, Stanford Research Systems), with the modulation signal being supplied by a lock-in amplifier (SR830, Stanford Research Systems). A set of modulation frequencies in the range 50–100 kHz was used. The laser light was filtered prior to illuminating the sample using an 800 nm short-pass (OD 4) filter to remove undesired side-modes.

Light is scattered by or emitted from the sample, which was collected and collimated with the use of a parabolic mirror. The collimated light was routed through an automated filter slider mounted with two sets of filters: 800 nm short-pass (OD 4) or 800–875 nm bandpass (OD 8). These filters allow for differentiation between the laser light and light from the $ZnP_2$ band-to-band transition. The filtered light was then focused with a condenser lens onto an avalanche photodiode detector (APD130A, Thorlabs), which routes the voltage signal to the lock-in amplifier. The amplifier, using its own oscillator frequency as the reference, was used to measure the phase difference between the signal from the scattered laser light and that of the sample photoluminescence. This phase difference ($\varphi$) was then used to estimate the phase lifetime ($\tau$):

$$\tau = \tan(\varphi)/\omega, \quad (5)$$

where $\omega = 2\pi f$ with $f$ the modulation frequency. This expression arises from analyzing the fluorescence emission intensity of an exponentially decaying system:[71]

$$I(t) = I_0 \exp(-t/\tau), \quad (6)$$

to sinusoidally modulated light:

$$L(t) = a + b * \sin(\omega t), \quad (7)$$

with $b/a$ the modulation depth (20% in our measurements). $I(t)$ at any time is proportional to the excess carrier concentration:

$$N(t) = A + B * \sin(\omega t - \varphi). \quad (8)$$

This equation suggests that the system responds with the same frequency, albeit with a phase shift and different modulation depth. For an exponentially decaying system, the differential equation describing the time dependence is:

$$\frac{dI(t)}{dt} = -\frac{I(t)}{\tau} + L(t). \quad (9)$$

Solving this equation for the relevant conditions yields Equation 5. A detailed derivation can be found in Ref. 71.

**DC photoconductive current measurements**

Ohmic contacts were established by pressing indium pads on a $ZnP_2$ crystal. To measure the photoconductive current, the sample was illuminated with a 780 nm laser diode modulated at 227 Hz (square wave) using an SRS DS345 function generator. Positioned 1 cm above the sample, the laser diode beam divergence ($8^\circ \times 30^\circ$) produced an elliptical spot size of 6.34 $mm^2$. With a maximum output power of 10.5 mW, the resulting power density at the sample surface was 0.164 $W/cm^2$ (determined by measuring the power at the source of the light using a power meter as well as the area covered, i.e., the spot size). The resulting current was conditioned using an SR570 low-noise current preamplifier and a low-pass filter. An SR830 lock-in amplifier was utilized to isolate the signal at 227 Hz. For photoconductive current decay measurements, the modulation frequency was adjusted to 111 kHz, and the lock-in amplifier was replaced with an oscilloscope to resolve the current decay during the laser-off cycle.

**PLQY measurements**

PL quantum yield (PLQY) measurements were performed using a home-built setup implementing a variation of the de Mello method,[76] The light source used was 638 laser light with 100% sinusoidal modulation, with the peak current producing illumination corresponding to approximately 1 sun equivalent photon flux. The laser system was the same as the one used in the phase fluorometry measurements. A modulation frequency of 993 Hz was used. The laser light was filtered prior to illuminating the sample using an 800 nm short-pass (OD 4) filter to remove undesired side-modes.

The setup utilizes a $BaSO_4$ coated integrating sphere (Newport, 819D Spectraflect Coated). The sample is mounted on a home-built, stepper-driven, carriage constructed from polished aluminum and PTFE (Teflon, McMaster-Carr). The carriage moves the sample into the laser beam, out of the laser beam but in the sphere, and out of the sphere altogether, as per the de Mello method.

Light from the integrating sphere escapes via a pinhole and is collimated with the use of a plano-convex lens. The collimated light was routed through an automated filter slider mounted with two sets of filters: 800 nm short-pass (OD 4) plus two OD1 absorptive filters, or 800–875 nm bandpass (OD 8). These filters allow for differentiation between the laser light and light from the $ZnP_2$ band-to-band transition. The filtered light was then focused with a condenser lens onto an avalanche photodiode detector (APD130A, Thorlabs), which routes the voltage signal to the lock-in amplifier. The amplifier, using its own oscillator frequency as the reference, measures the voltage amplitude for the laser and photoluminescent light for each of the three stepper positions. These values are corrected to account for the detector responsivity as well as the laser light attenuation by the absorptive filters. These corrected values are treated as the integrated signals and used to calculate the reported PLQY value.

### DC Hall effect measurements

Hall effect measurements were conducted using a Quantum Design Physical Property Measurement System (QD-PPMS) Evercool II, equipped with the Electrical Transport Option (ETO). A roughly rectangular-shaped crystal was used to measure Hall resistance with a 5-probe setup. Electrical contacts were made using platinum wires attached with conductive silver paste. The distance between the voltage probes is 0.52 mm. The Hall voltage was calculated using the equation $V = I \times R$ (where $I = 0.0001$ A). Hall coefficient ($R_H$) was determined from the slope of the Hall voltage versus magnetic field plots at different temperatures. The measurements were carried out for a range of magnetic fields from $-3$ to $3$ T. The Hall carrier concentration ($n$) and mobility ($\mu$) were calculated using $n = 1/|R_H * e|$, where $e$ is the electron charge, and $\mu = |R_H/r|$, where $r$ is the electrical resistivity.

## ACKNOWLEDGMENTS

This work was supported by the U.S. Department of Energy, Office of Science, Basic Energy Sciences, Division of Materials Science and Engineering, Physical Behavior of Materials program under award number DE-SC0023509 to Dartmouth and was authored in part by the National Laboratory of the Rockies, operated by The Alliance for Advanced Energy, LLC, for the U.S. Department of Energy (DOE) under contract no. DE-AC36-08GO28308. All computations, syntheses, and characterizations were supported by this award unless specifically stated otherwise. This research used resources of the National Energy Research Scientific Computing Center (NERSC), a DOE Office of Science User Facility supported by the Office of Science of the U.S. Department of Energy under contract no. DE-AC02-05CH11231 using NERSC award BES-ERCAP0023830. Funding for microwave conductivity measurements and analysis on semiconductor powders is provided by the Materials Chemistry Program, Materials Sciences and Engineering Division, Office of Basic Energy Sciences, U.S. Department of Energy under grant DE-SC0023316. The PPMS instrument used for Hall effect measurements was supported by Ames National Laboratory, U.S. Department of Energy, which operates under the contract DE-AC02-07CH11358. We thank Prof. J. V. Zaikina (Iowa State University) for access to the UV/Vis/NIR spectrophotometer. The views expressed in the article do not necessarily represent the views of the DOE or the U.S. Government.

## AUTHOR CONTRIBUTIONS

Z.Y. and G.H. worked on the first-principles calculations, including the high-throughput computational screening. Z.Y., K. K., J. L. and G. H. conceived the idea of investigating the zinc polyphosphide. G.A., S.G., G.V., J.T.R., M.R.H., S.D., Y.C., S.R.B., J.L. and K.K. worked on the synthesis, structural, chemical, and stability characterization. S.Q., S.G., G.L.E., G.K., O.G.R., J.R.P., S.D., D.P.F., S.R.B. and J.L. worked on the optoelectronic characterization. Z.Y. and G.H. coordinated the computations and experiments, and prepared the manuscript with inputs from all co-authors. A.Z., D.P.F., S.R.B., K.K., J.L. and G.H. supervised the project.

## DATA AVAILABILITY

The data supporting the findings of this study are provided in the article and its Supplementary Information and are available from the corresponding author upon request.

## DECLARATION OF INTERESTS

The authors declare no competing interests.